\documentclass[final,1p,times]{elsarticle} 
\usepackage{graphicx} 
\usepackage{amssymb} 
\usepackage{amsthm} 
\usepackage{lineno} 

\journal{Nuclear Physics A} 
\begin{document} 

\begin{frontmatter} 
\title{Muon Production in Relativistic Cosmic-Ray Interactions}
\author{Spencer R. Klein}
\address{Nuclear Science Division, LBNL, Berkeley, CA, 94720 USA and the Physics Department, University of California, Berkeley, 94720 USA}

\begin{abstract} 

Cosmic-rays with energies up to $3\times10^{20}$ eV have been observed. The nuclear composition of these cosmic rays is unknown but if the incident nuclei are protons then the corresponding center of mass energy is $\sqrt{s_{nn}} = 700$ TeV.   High energy muons can be used to probe the composition of these incident nuclei.
The energy spectra of high-energy ($>$ 1 TeV) cosmic ray induced muons  have been measured with deep underground or under-ice detectors.  These muons come from pion and kaon decays and from charm production in the atmosphere.  

Terrestrial experiments are most sensitive to far-forward muons so the production rates are sensitive to high-$x$ partons in the incident nucleus and low-$x$ partons in the nitrogen/oxygen targets.  Muon measurements can complement the central-particle data collected at colliders.  This paper will review  muon production data and discuss some non-perturbative (soft) models that have been used to interpret  the data.  I will show measurements of TeV muon transverse momentum ($p_T$) spectra in cosmic-ray air showers from MACRO, and describe how the IceCube neutrino observatory and the proposed Km3Net detector will extend these measurements to a higher $p_T$ region where perturbative QCD should apply.  With a 1 km$^2$ surface area, the full IceCube detector should observe hundreds of muons/year with $p_T$ in the pQCD regime.  

\end{abstract} 

\end{frontmatter} 


\section{Introduction}

Surface air shower arrays and air-fluorescence detectors have observed cosmic-ray showers with energies up to $3\times10^{20}$ eV.  However, despite decades of effort (the first $10^{20}$ eV showers were seen in the 1960's \cite{volcano}), we still know very little about these particles. 

At energies below $10^{15}$ eV, direct measurements by balloon and satellite detectors have shown that cosmic rays include nuclei, from hydrogen to iron with a sprinkling of heavier elements. This is consistent with the composition that is expected from supernovae \cite{supernova}, a  likely source for lower energy cosmic rays.    However, at higher energies the composition is poorly known. 

Higher energy fluxes are too low for direct measurements so we must use indirect approaches, with much larger terrestrial detectors that observe the shower particles created when the cosmic-ray interacts in the atmosphere.  The atmosphere is 28 radiation lengths (11 hadronic interaction lengths) thick, so measuring the cosmic-ray composition is akin to doing particle identification from the back of a calorimeter.  Detectors measure the total energy, not the energy per particle.  The challenge is to distinguish a $10^{17}$ eV proton (for example) from a $10^{17}$ eV (total) iron nucleus, with $A=56$.   As $A$ rises,  the initial energy is spread among more particles and the shower develops faster, reaching its maximum particle count higher in the atmosphere.

Shower development has been studied by air fluorescence detectors, which observe nitrogen fluorescence induced by charged particles moving through the atmosphere. These measurements are calorimetric, so they provide a fairly direct measure of the total shower energy of the shower. One can also observe the shower development as it progresses through the atmosphere, and thereby probe the composition.

Shower particles that reach the surface are studied with  air shower arrays, grids of scintillator or water/ice Cherenkov sampling detectors which sample the electrons and photons which reach the ground. These arrays can cover up to 3000 km$^2$. They measure the shower arrival direction (using timing),  energy (via particle density measurements), and lateral spread, with the lateral spread being somewhat sensitive to composition.   Co-located muon detectors measure muons from pion and kaon decay.  Most of these muons have energies of a few GeV.   The ratio of muons to electromagnetic energy is also composition sensitive, with heavier nuclei producing more muons.  Unfortunately, the inferred energy and composition are dependent on the hadronic interaction model used to simulate the showers.  

\section{TeV Muons}

The LEP experiments have used magnetic spectrometers to study cosmic-ray muons with energies up 3 TeV/c \cite{LEP}.   TeV muons can also be studied by putting a detector underneath a kilometer or more of rock or ice.  These muons are produced early in the shower, and so are somewhat more directly sensitive to the initial ion composition.  However, a very large detector, ($10^3$ to $10^6$ m$^2$) is needed to get good statistics. For example, the 72 m by 12 m MACRO experiment in the Gran Sasso Laboratory lay below 1500 meters of water equivalent (mwe) of rock shielding \cite{MACRO}.  

These experiments have measured the muon energy spectra and multiplicities, and also the lateral separation (decoherence).  These distributions are compared with models with varying cosmic-ray compositions.  Unfortunately, the choice of hadronic interaction model can have a significant affect on the inferred composition.  MACRO also took data in coincidence with the EASTOP air shower array.  More recently, the SPASE air shower array at the South Pole and AMANDA in-ice detector took data in coincidence.  These combined datasets have reduced, but still significant sensitivity to the choice of hadronic model.  Because these muons are predominantly produced from low $p_T$ pions and kaons, these models are necessarily quite phenomenological \cite{models}. 

\begin{figure}[t]
\centering
\includegraphics[scale=0.3]{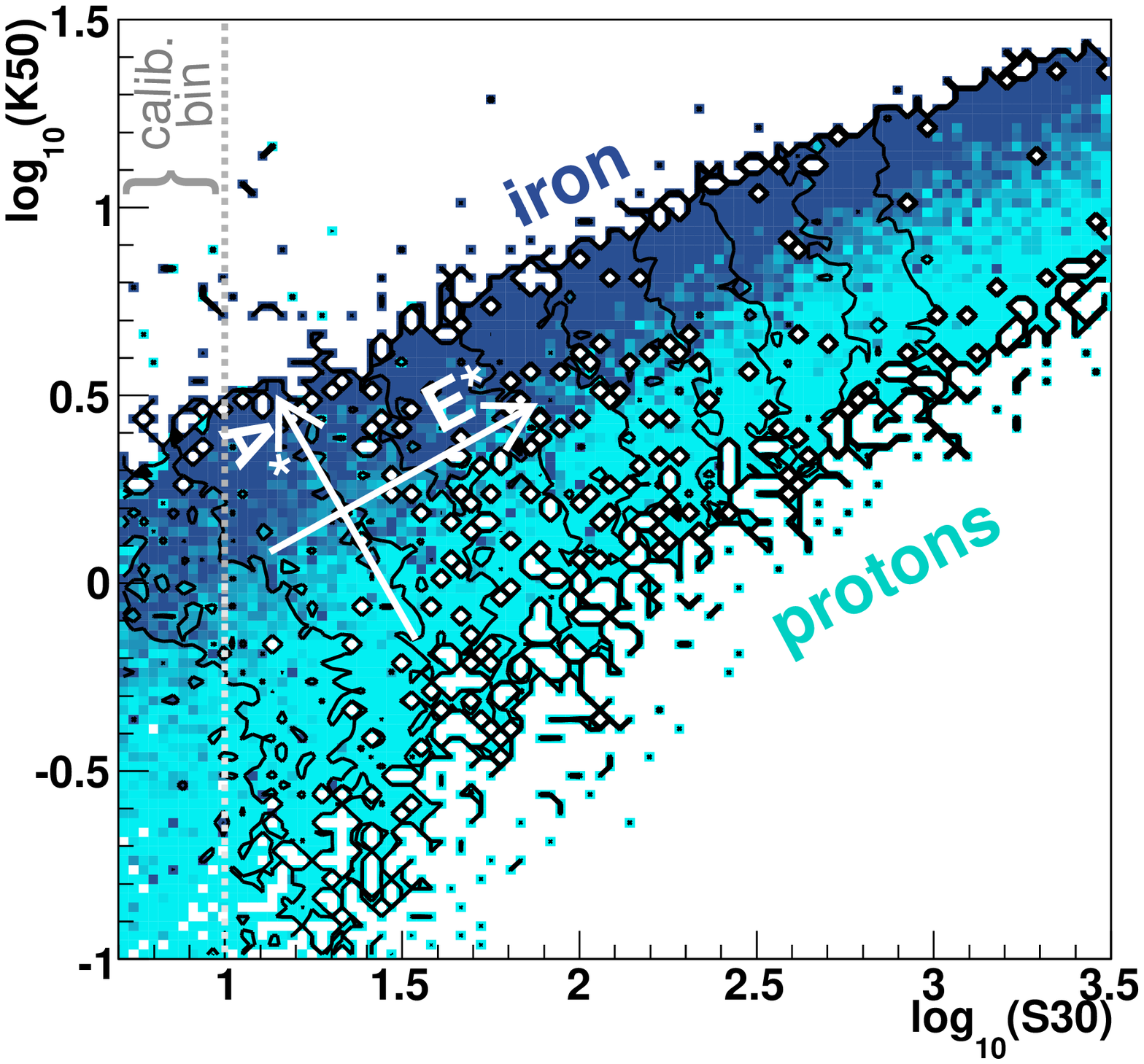}
\includegraphics[scale=0.29]{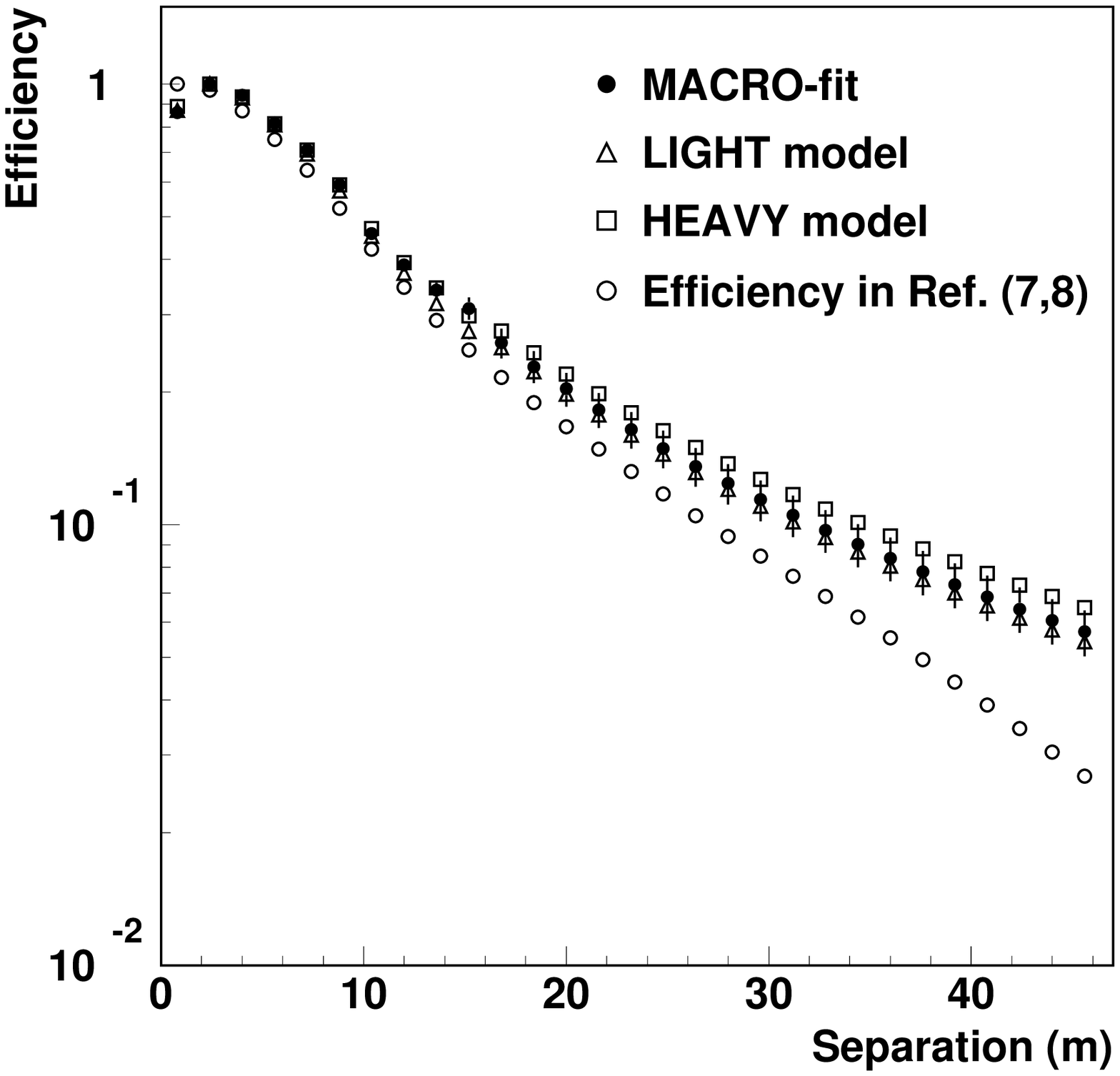}
\vskip -0.1 in 
\caption[]{(left) Combined data from the SPASE-II south polar array and the AMANDA underground muon detector.  S30 is a measure of the surface energy, while K50 is a measure of the light produced by muons in the underground array.   Iron showers produce more muons.  The data prefers a mixed composition \cite{SPASEICRC}.
(right)  The muon decoherence (separation) function measured by MACRO, from 0 to 50 m separation, compared with the results of a light and a heavy composition model.  The data is intermediate between the two models \cite{MACRO}. }
\label{fig:AMANDA}
\end{figure}

Figure \ref{fig:AMANDA} shows the SPASE/AMANDA composition measurement \cite{SPASEICRC}.  The x axis shows the surface energy measured by SPASE, while the the y axis shows the aggregate muon energy in AMANDA.  The blue shading shows the expectation for all-proton and all-iron simulations.  The data (points) are in-between, indicating an intermediate composition.  

\section{High $p_T$ Muon in Air Showers}

\begin{figure}[t]
\centering
\includegraphics[scale=0.32]{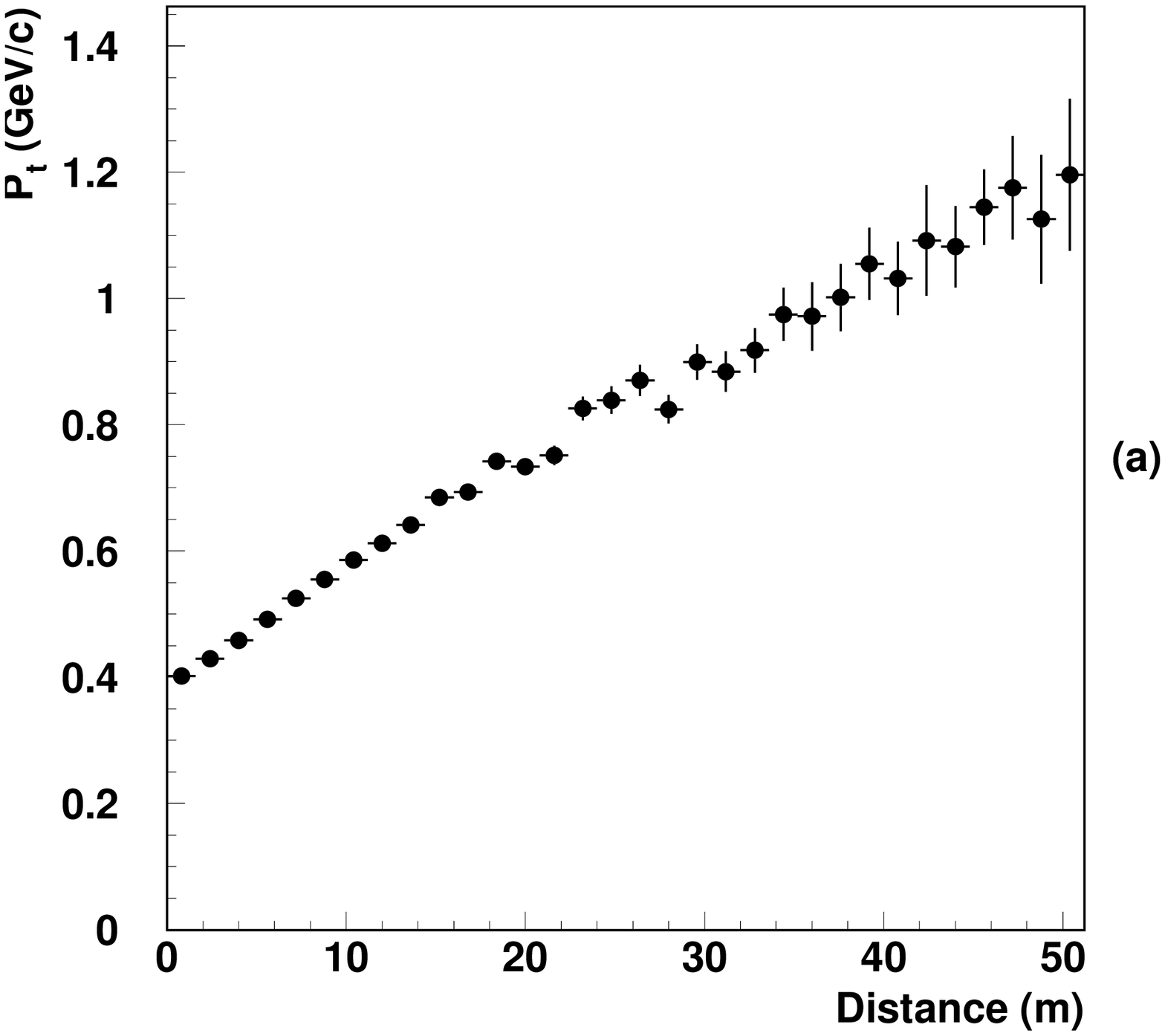}
\hskip 0.7 in
\includegraphics[scale=0.4]{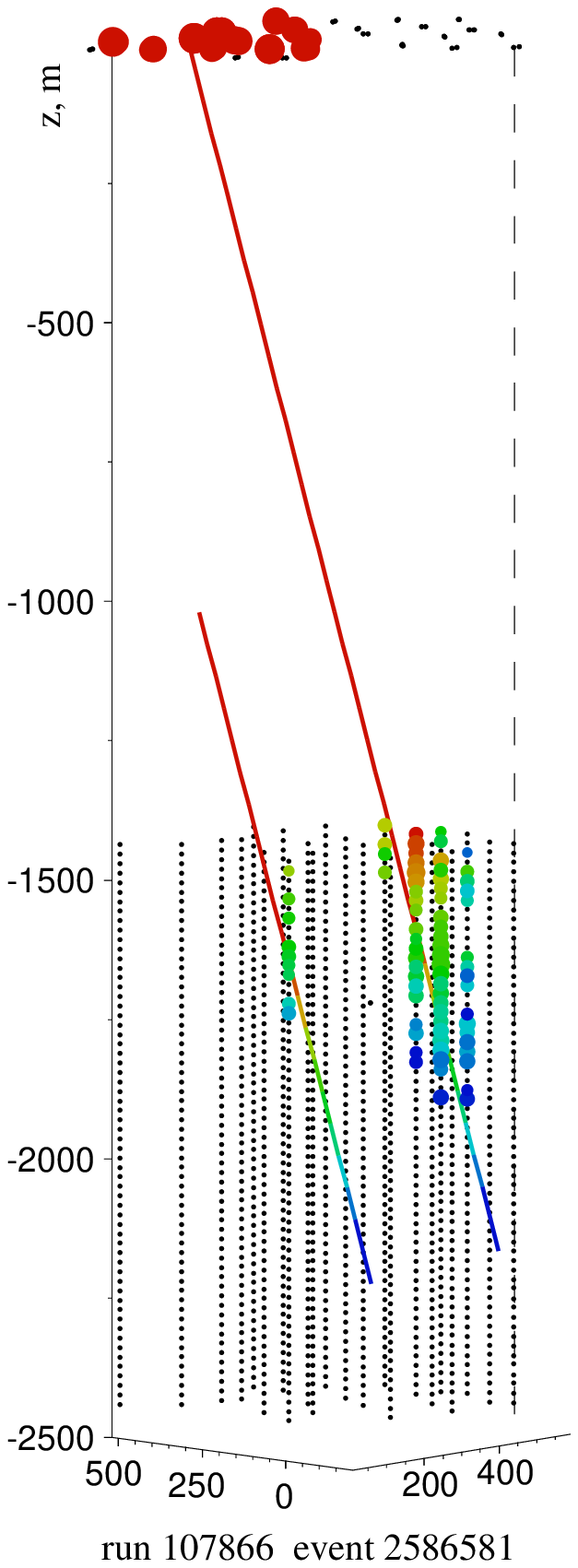}
\caption[]{(left) The relationship between $p_T$ and separation determined by MACRO \cite{MACRO}.
(right)   An interesting event observed by IceCube.  An air shower hits the 11 stations in the IceTop surface array, and light is seen in 96 buried optical modules. 84 of the modules are on 4 strings near the extrapolated air shower core. The remaining 12 modules are on another string about 400 meters from the projection \cite{IceCubeICRC}. }
\label{fig:IceCubeevent}
\end{figure}

High $p_T$ muons will allow perturbative QCD (pQCD) to be used to interpret the data and infer composition.  Since TeV muons are created high in the atmosphere, the high $p_T$ muons are isolated from the bundle of predominantly lower $p_T$ muons.  The lateral separation $d$ from the shower core at the Earths surface is
\begin{equation}
d = h \frac{p_T}{E_\mu}
\end{equation}
where $h$ is the interaction height and $E_\mu$ is the muon energy.  For a typical interaction height of 25 km, a 500 GeV muon with $p_T=4$ GeV/c will be 200 m from the shower core.  The MACRO  collaboration has measured muon pair decoherence (separations) out to 50 meters, as in Fig. \ref{fig:AMANDA} (right), and compared them to different models.  
They also used detailed simulations, shown in Fig. \ref{fig:IceCubeevent} (left) to relate separation to mean $p_T$.
In MACRO, 50 m corresponds to a mean $p_T$ of 1.2 GeV, not fully in the pQCD regime. 

With a 1 km$^2$ surface area, the IceCube detector and IceTop surface array will be able to measure muons out to much larger separation distances, where pQCD holds robustly.  Rate estimates for charmed quarks and for pions and kaons in jets depend on the minimum accessible $E_\mu$ and $d$, but the complete IceTop/IceCube should observe hundreds of high $p_T$ muons each year.  Most of these muons are from charmed particle decays, with a smaller number from pions and kaons in jets. The composition has a large effect on the muon $p_T$ spectra, especially near the kinematic limits.  There are many more $10^{16}$ eV partons in a $10^{17}$ eV proton than in a $10^{17}$ eV iron nucleus, at any $Q^2$.  

Figure \ref{fig:IceCubeevent} (right) shows one event found by IceCube, showing an isolated muon about 400 m from the shower core.  The collaboration has also observed near-horizontal muon pairs. These muons have a initial energies of at least several TeV \cite{Lisa}.  The proposed Km3Net detector could make similar studies. 

The measured muon decoherence spectrum can be compared with predictions based on pQCD and different composition models.  In the longer run, one can also use this data to probe shadowing at low Bjorken$-x$.  Because of the fixed-target geometry, the observed muons are produced in the far-forward region, and come from the collision of a high$-x$ parton in the incident particle with a low-$x$ parton in the $N_2$ or $O_2$ target.  The data will therefore be sensitive to shadowing in nitrogen or oxygen.  The accessible kinematic range will depend on the minimum observable muon-core distance, but it may reach below $x=10^{-5}$ \cite{ISVHECRI}.  

\section{Conclusions}

TeV muons can be used to probe cosmic-ray interactions at energies above 1 PeV, where, despite decades of effort we do not understand the incident cosmic-ray composition.  The IceCube detector, now being built at the South Pole, will be large enough to study high $p_T$ muon production in air showers.  This offers the possibility to study the composition in the context of a pQCD model.  It also offers nuclear physicists the opportunity to study production of far forward muons, potentially probing nuclear shadowing at very small $x$.  

\smallskip

{\bf Acknowledgments.} I thank my IceCube and STAR Collaborators for useful comments on this work.   Lisa Gerhardt has contributed greatly to this study of high $p_T$ muons.
This work was supported in part by the National Science Foundation under Grant 0653266 and the Department of Energy under Contract DE-AC-02-05CH11231.  

\end{document}